\documentclass[manuscript,screen]{acmart}
\AtBeginDocument{%
  }

\setcopyright{acmlicensed}
\copyrightyear{2025}
\acmYear{2025}
\acmDOI{XXXXXXX.XXXXXXX}

\acmISBN{978-1-4503-XXXX-X/2018/06}

\begin{document}

\title{Are Quantum Voting Protocols Practical?}

\author{Nitin Jha}
\authornote{Both authors contributed equally to this research.}
\email{njha1@students.kennesaw.edu}
\author{Abhishek Parakh}
\authornotemark[1]
\email{aparakh@kennesaw.edu}
\affiliation{%
  \institution{Kennesaw State University}
  \city{Marietta}
  \state{GA}
  \country{USA}
}

\renewcommand{\shortauthors}{Jha and Parakh 2025}

\begin{abstract}
Quantum voting protocols aim to offer ballot secrecy and publicly verifiable tallies using physical guarantees from quantum mechanics, rather than relying solely on computational hardness. This article surveys whether such quantum voting protocols are practical. We begin by outlining core mathematical ideas such as the superposition principle, the no-cloning theorem, and quantum entanglement. We then define a common system and threat model, identifying key actors, trust assumptions, and security goals. Representative protocol families are reviewed, including entanglement-based schemes with central tallying, self-tallying designs that enable public verification, and authority-minimized approaches that certify untrusted devices through observable correlations. Finally, we evaluate implementation challenges, including loss, noise, device imperfections, scalability, and coercion resistance, and discuss realistic near-term deployment scenarios for small-scale elections.

\end{abstract}


\maketitle

\section{Introduction}
Like Delphi’s oracle, elections must be beyond reproach. Election protocols, whether electronic or paper ballots, must satisfy two competing security guarantees - maintain the secrecy of ballots, and provide publicly verifiable results. Classical cryptography-based solutions solve this by using end-to-end encryption, scrambling mechanisms called mixed-nets, public bulletin boards, and zero-knowledge proofs. Some of these systems have even seen deployments at smaller scales \cite{centrone2022quantum, kumar2022experimental}, mostly in Europe, while the most popular deployment today remains the electronic voting machines with voter-verified paper trails (VVPT) \cite{ansari2008evaluating, hughes2022usability}. 

The security of classical cryptographic algorithms that support electronic voting protocols relies on the mathematical complexity, and therefore, long-term storage of election data and audit trails can be an issue since an attacker who cannot break today’s encryption can still capture traffic and decrypt it later. While the adoption of post-quantum cryptography (resistant to quantum attacks) offers some promises, it still relies on unproven hardness assumptions \cite{mmaduekwe2024cybersecurity}.

Quantum voting claims to provide an alternative solution to secure electronic voting based on the fundamental principles of quantum mechanics. For example, the no-cloning principle prevents copying of unknown states, allowing any tampering to be detected. Entanglement allows parties to share correlations that reveal collective properties while hiding individual contributions, helping prevent deanonymisation of individual inputs. Consequently, a quantum voting protocol can be designed so that each voter encodes a choice into a local operation (such as quantum bit rotation) on part of a shared quantum resource, while only joint measurements or public post-processing reveal the overall tally. Probing any single piece of quantum information should not leak meaningful information, and tampering attempts are revealed as noise or as inconsistent correlations.

A typical electronic voting protocol consists of the following parties and rules: (1) Voters: each voter holds a credential and applies a local operation (and, in some variants, publishes a small classical share), (2) A registrar (eligibility authority) issues one credential per eligible voter and publishes revocation or eligibility lists. It is trusted for the correctness of issuance, but need not be trusted for tally privacy. (3) A quantum source prepares and distributes ballot states or multipartite entanglement. In authority-minimized designs, the source can be untrusted and certified via publicly verifiable test rounds. (4) A collector (central-tally variants only) performs a joint measurement and posts outcomes with audit data. Privacy should not rely on trusting the collector, while correctness is enforced by evidence on a public bulletin board (append-only, publicly readable), and (5) Public verifiers or auditors recompute checks from the posted evidence. Finally, the minimal trust profile is defined in the following way. Classical communication channels must be authenticated to prevent tampering or impersonation. The registrar is trusted only for issuing one credential per voter and preventing double voting, but not for ballot privacy. Thus, a voting pilot is a limited, real-world trial of the protocol, not a system role.

The idea is not to develop one universal quantum voting mechanism but rather a family of these protocols depending on different trade-offs in trust, hardware, and usability. Early work treated voting as a special case of privacy-preserving multiparty computation performed with entangled quantum states. However, later proposals developed self-tallying elections, where anyone can compute the result from public data without relying on a designated vote counter. Some newer designs even try to remove trusted middlemen (the collector or the counter). They use a shared quantum state made by an untrusted source. If the measurement results show the expected pattern, we know things were done correctly even without a designated authority. Another, more practical, approach avoids storing qubits as today's quantum memories have limited capacities. It uses optical signals that are prepared, sent, and measured immediately, which lowers the difficulty of building smaller prototypes on short links.

The use of quantum mechanics principles to achieve secure voting prototypes comes with its own challenges. The theoretical security guarantees are often hard to implement, as physical devices are imperfect: light signals fade and pick up noise in the channel, and parts like detectors or modulators can leak clues through side channels, for example, small timing delays, power spikes, optical reflections, radio noise, or logging artifacts that line up with a voter’s action. By calibration, we mean keeping all sites lined up so the source, modulators, interferometers, and detectors operate within set ranges: clocks stay in sync, optical phase and polarization stay stable, power levels and detector thresholds stay within bounds. Apart from this, theoretical security cannot guarantee societal goals. Coercion resistance, for example, means a voter cannot prove their vote to the adversary to avoid compliance by a voter to the adversary's demands \cite{juels2005coercion}. Therefore, it's important to clarify where quantum properties can strengthen privacy and verifiability, where they cannot, and how those insights fit into broader election practices.

This article focuses on the properties of voting principles: privacy and anonymity, eligibility and one-person-one-vote, fairness, public verifiability, and receipt-freeness with coercion resistance. Quantum protocols achieve these to different degrees and under strict assumptions. In the near term, realistic use often means small networks on short and well-controlled links, where self-tallying and memory-free optical designs use available hardware. City or national elections would require reliable multipartite entanglement or repeater-assisted links, which are open questions for many active research studies. We provide a brief survey of different classes of quantum voting protocols, the advantages of using them, and their practicality of deployment.

The article is structured as follows: Sec \ref{Sec:basics} goes over the basic quantum mechanics principles that are central to the security guarantees of any quantum algorithm, and it also covers the basic theory and elements associated with quantum voting schemes. Sec \ref{Sec:system} specifies the system architecture and threat model, including roles, trust assumptions, and security goals. Sec. \ref{Sec:family} surveys representative voting protocols, grouped into entanglement-based, self-tallying, authority-minimized, and memory-free optical designs. Sec. \ref{Sec:adv} analyzes advantages and limitations, with attention to scalability, device imperfections, and coercion resistance. Sec. \ref{Sec:challenges} evaluates practicality and near-term deployment patterns, and outlines integration with classical infrastructure for pilots on short, well-controlled links. It also summarizes open problems and engineering challenges. Sec \ref{Sec:Conclusion} concludes the work and the several topics covered across the article.

\section{Quantum Basics for Voting}
\label{Sec:basics}
Before jumping into the details of quantum voting protocols, we need to understand some of the fundamental properties that guarantee their security. To put it briefly, laws of quantum mechanics restrict the nature of information that can be copied and measured by any party present in the system (both legitimate and illegitimate ones). This further allows for carefully constructed correlations, which can help in achieving a final voting outcome while keeping individual identities private. 

\begin{enumerate}
    \item[1.] \textbf{No-Cloning Principle}: Classical information can be copied, even when it is encrypted, and stored for later use by the adversaries. However, the fundamental laws of quantum mechanics do not allow anyone to make a copy of an unknown quantum state. Therefore, quantum computers have to perform blind operations (i.e., letting a quantum state evolve under different operators without knowing the exact state). Fig. \ref{fig:no-cloning} presents a schematic diagram representing the no-cloning principle.

    \begin{figure}[h!]
        \centering
        \includegraphics[width=0.7\linewidth]{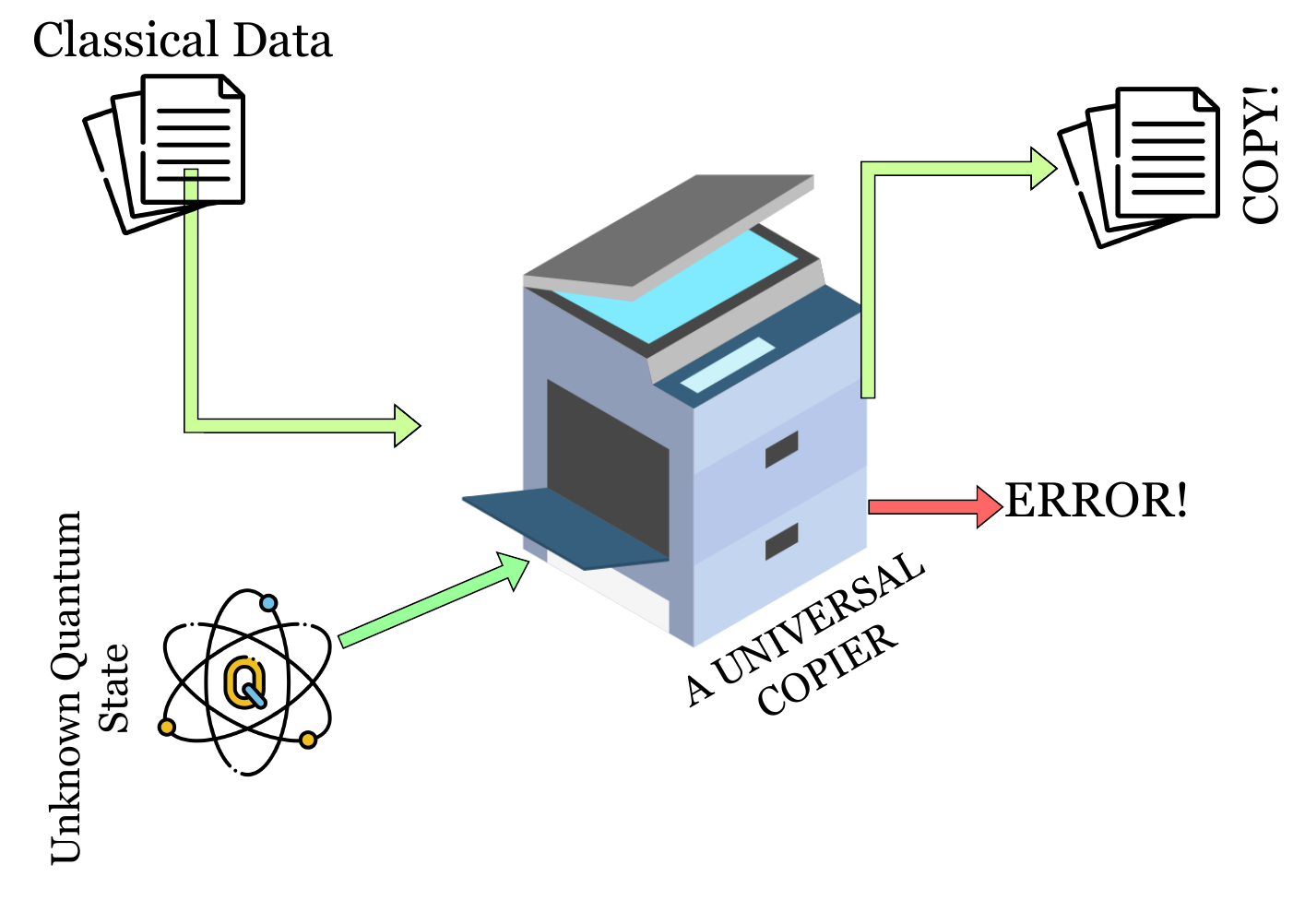}
        \caption{Schematic representation of the no-cloning principle. We assume there is a universal copier machine; we feed into it any classical data (encrypted or otherwise) it successfully returns a copy of it. However, when we try copying a quantum state (which is unknown to us, i.e., not prepared by us), it is unsuccessful. This demonstrates that the no-cloning principle disallows us from copying an arbitrary unknown quantum state.}
        \label{fig:no-cloning}
    \end{figure}

    \item[2.] \textbf{Quantum Entanglement}: This is one of the mysterious quantum properties that lacks a classical counterpart for simple explanations. However, it can be understood as correlations between two qubits that cause them to behave in the same fashion, i.e., the result of an operation on one of the qubits from an \textit{entangled pair} will show on the other qubit as well, despite the physical distance between them!

    \item[3.] \textbf{Quantum Superposition}: All classical information is in a deterministic state, i.e., the exact nature of the state can be found out. However, qubits exist as a combination of all possible outcomes, and only when one measures them does a deterministic state appear. This property is known as \textit{quantum superposition}, and it allows quantum computers to perform parallel computations on multiple states \cite{markidis2024quantum}.

    \item[4.] \textbf{Local Ballot Encoding}: Each voter records a choice by making a small, local change on their share of a prepared quantum state. This change can be a simple quantum gate operation or a tiny phase shift (also called discrete–variable protocols). While in optical protocols, it can be a small rotation of the phase of a light pulse that is measured right away. The key idea is \textit{locality} - the action happens only in the voter’s end, so the choice is not revealed to others, and no extra copy of the ballot is created.

    \item[5.] \textbf{Vote tally}: There are two simple ways to see the result. First, a central device gathers all the qubits, makes one joint measurement, and then publishes the outcome together with basic check data to show that the run was executed well. Second, in a self-tallying mode, each voter will post a small classical value computed from their local operation, and any observer can combine these public values to get the final count. The self-tallying option reduces dependency on a central counter, but it usually requires tighter management of how the initial quantum state was prepared and distributed among the voters. 
\end{enumerate}

\section{System and Threat Models}
\label{Sec:system}
Now that the basic principles are discussed, this section defines several actors, channels, evidence, and adversaries assumed in the design of any quantum voting principle. These are general definitions that can help understand a common frame of reference in which different schemes can be compared on privacy, verifiability, fairness, and practicality.

\subsection{Different Actors}
\begin{enumerate}
    \item \textbf{Voters} encode choices through local operations on provided quantum states and publish limited classical data as specified by the protocol.
    \item \textbf{Registrar} (or eligibility service) issues credentials that bind a real person to a single right to vote.  Eligibility is only required for one-person-one-vote and not necessarily for the tally itself. 

    \item \textbf{State source} prepares and distributes ballot states or multipartite entanglement. In authority-minimized variants, this source may be untrusted, with correctness inferred from observable correlations and test rounds.

    \item \textbf{Collector} performs a joint measurement when a protocol uses central tallying. Self-tallying designs don't require this role.

    \item A \textbf{Public bulletin board} is an append-only, publicly readable log where ciphertexts, commitments, or compact shares are posted together with proofs and audit data.

    \item \textbf{Public verifiers} are any observers who recompute checks from the posted evidence. These roles match classical end-to-end verifiable (E2E) systems, with the state source and quantum channels added for the voting context.
    
\end{enumerate}

\subsection{Channels and randomness}

Each run uses a quantum channel to distribute ballot-bearing states and an authenticated classical channel to coordinate timing, publish different parameters, and record audit data on a public bulletin board. Authentication prevents man-in-the-middle changes to announcements and device settings and is a standard practice in quantum cryptography. A commitment is a short string posted to the bulletin board that binds a party to a hidden value until a later open step; opening means revealing the value together with its decommitment data so any verifier can recompute the commitment and confirm it matches the original posting. Think of it as a sealed envelope: binding means it cannot be changed after posting, hiding means observers learn nothing before it is opened. Initial authentication keys can be issued by the registrar during enrollment or derived from pre-shared secrets, then rotated using fresh material from subsequent runs or standard key-agreement procedures.

\subsection{Evidence and checks}
Using the public evidence, anyone should be able to confirm the following: (i) ballots were well-formed, i.e., it obeys the election’s rules and the protocol’s encoding rules, (ii) each ballot is recorded exactly as the voter cast it, and the resulting collection is the recorded set, (iii) the tally is calculated from the recorded set, and (iv) protocol steps that protect the integrity of the cast ballots were followed. 

Quantum schemes add parameter-estimation statistics that bound disturbance during state distribution and ballot encoding. If observed errors are found to be higher than a fixed threshold value, the run is terminated, and it can be repeated.

\subsection{Security properties}
The following security properties must be satisfied: 

(a) \textit{privacy}, i.e., no link from a voter to their vote choice should be traced, 

(b) \textit{anonymity}, i.e., local shares reveal no vote information, 

(c) \textit{eligibility} and \textit{non-reusability}, i.e., only authorized voters, and at most one ballot each, 

(d) \textit{fairness} (no early partial results leak), and \textit{public verifiability} 

(e) \textit{Receipt-freeness and coercion resistance}, i.e., a voter cannot produce convincing proof of how they voted, and a protocol resists adversaries who instruct, monitor, or demand keys from voters. 

Quantum schemes provide security guarantees using quantum mechanics; however, social goals such as coercion resistance require careful protocol design. This, thus, becomes one of the important challenges towards designing scalable quantum voting protocols.

\section{Overview of Some of the Quantum Voting Protocols}
\label{Sec:family}
Now that we have discussed several components required to design a quantum voting protocol, we discuss some of the protocols proposed over the years. 

\subsection{Entanglement-based and central tally}
\label{Sec:Entangled}
\begin{figure}[h!]
    \centering
    \includegraphics[width=0.75\linewidth]{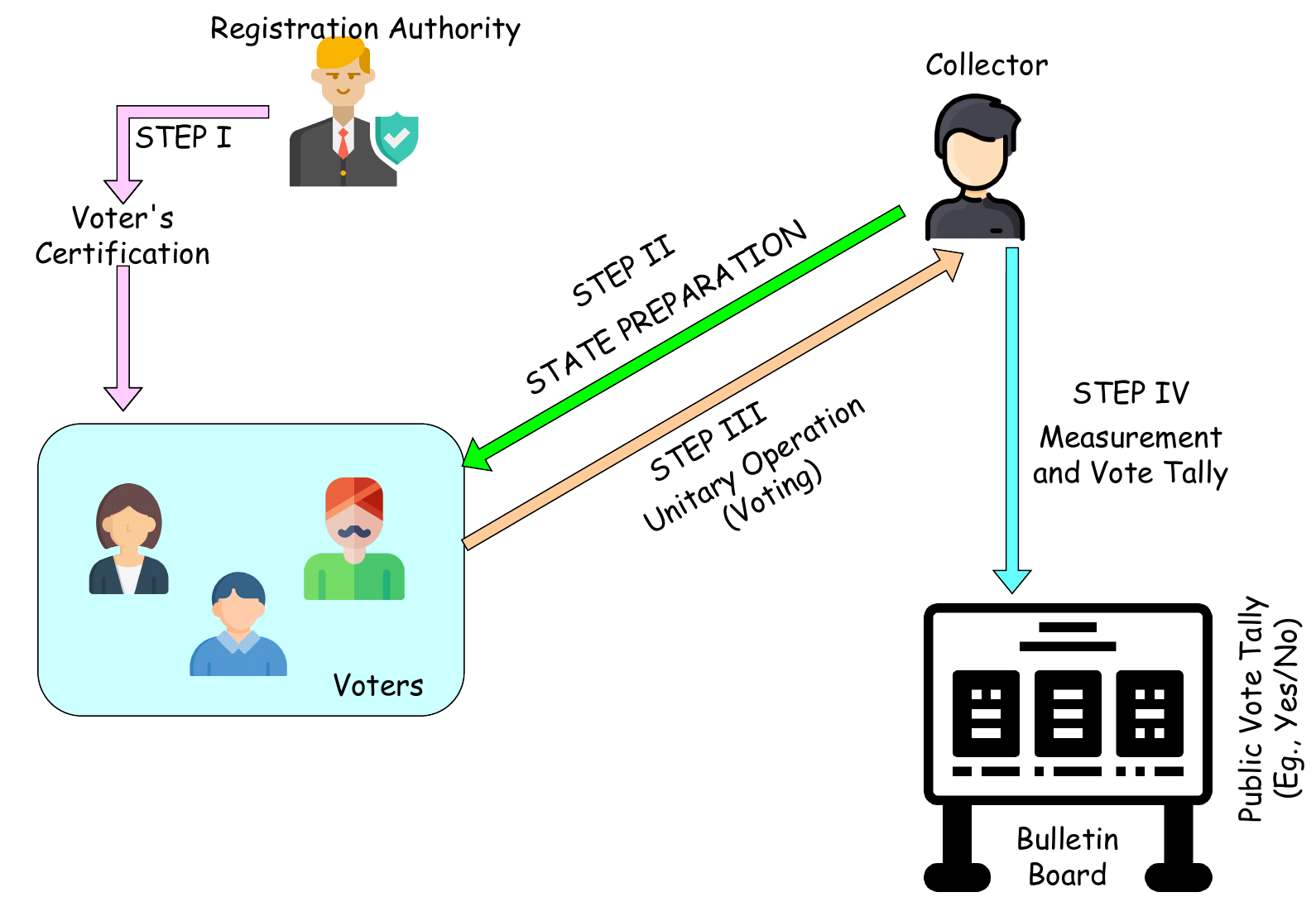}
    \caption{Schematic diagram which shows the four steps for the quantum voting scheme that uses a central collector. Step I is where a registration authority gives certification to all voters. Step II is where the collector prepares and distributes the multipartite quantum states to the voters. Step III is when each voter applies a local unitary operation (i.e., rotation) to register their vote and send the states back to the collector. Step IV is the final step where the collector measures the quantum states and posts the tally result and other test statistics so it's verifiable.}
    \label{fig:central}
\end{figure}

In this particular protocol, a central voting authority prepares a multipartite entangled quantum state and distributes one share of the prepared quantum state to each voter (as shown in Fig. \ref{fig:central}). A collector holds a joint-measurement device \cite{hillery2006towards, vaccaro2007quantum}. A joint-measurement device can be thought of as a box that takes all the quantum pieces back at the end and measures them together so that only the combined result is revealed. An authenticated bulletin board records evidence. The detailed steps are as follows:
\begin{enumerate}
    \item Distributes the entangled qubit shares and performs some initial test rounds to determine correlations and loss in the system.

    \item Each voter encodes a choice by using a unitary operation, i.e., for example, rotating the qubits by certain angles to denote \textit{yes} or \textit{no} (which is a phase flip for a yes/no).

    \item The collector performs a joint measurement on the shared states and thus gets the final tally. 

    \item The collector posts the outcome tally and different test statistics, which allows anyone to check the thresholds and recompute expected values.

\end{enumerate}
There are several weaknesses of using any entanglement-based system, especially from the implementation point of view. The loss associated with the channels and issues with misalignment of photon sources (in case of optical setups) can disturb the joint measurement. Furthermore, this works under the assumption of the collector being a trusted party.

\subsection{Self-Tallying Protocols}
\label{Sec:Self-tally}
In this protocol for quantum voting, voters act locally and then publish compact classical values derived from their operations. Any voter can combine these public values to compute the tally, thus reducing the dependency on a central collector and making public verification more convenient \cite{wang2016self}. The exact steps are as follows:
\begin{enumerate}
    \item Distributes the entangled qubit shares and performs some initial test rounds to determine correlations and loss in the system.

    \item Each voter, now, applies a local operation and derives a small classical share (for example, a phase value).

    \item Everyone posts the shares publicly on the bulletin board.

    \item Anyone can combine the classical outcomes from the public shares and get the final tally.
\end{enumerate}
\begin{figure}[h!]
    \centering
    \includegraphics[width=0.75\linewidth]{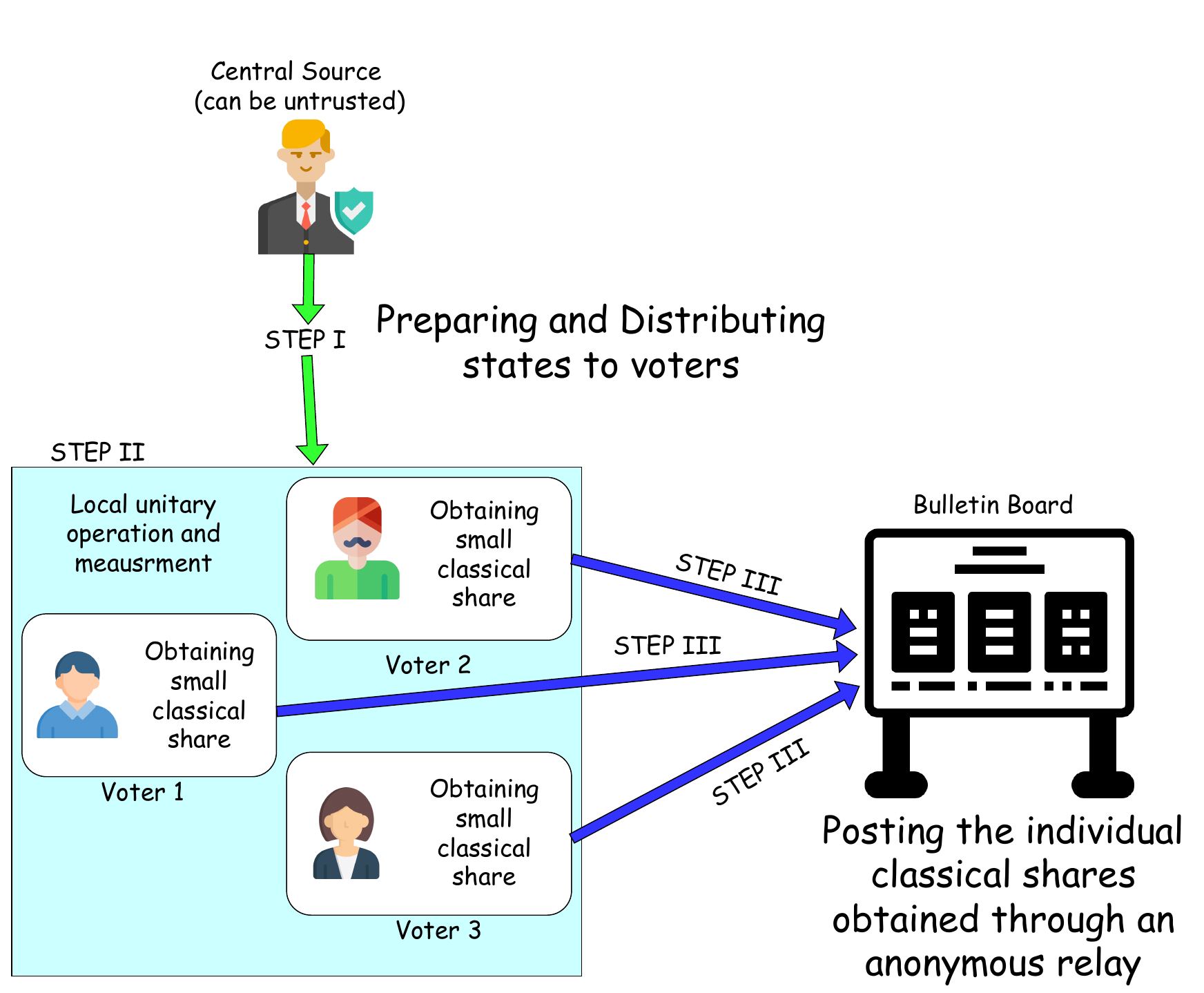}
    \caption{Schematic diagram which shows the mechanism of the self-tallying quantum voting scheme. Step I is where a central source distributes quantum shares to each voter; In Step II, each voter performs a local unitary operation and obtains a small classical share upon measurement; Step III is where each voter sends their obtained classical share (i.e., vote value) to the bulletin board using an anonymous relay to avoid deanonymisation.}
    \label{fig:selftally}
\end{figure}

Fig. \ref{fig:selftally} presents a schematic diagram representing various steps associated with self-tallying quantum voting schemes. By doing so, these schemes make the public verification of the vote easier than other methods. This scheme, however, works only under the assumption of honest distribution or strong correlation tests, and authenticated posted of their outcomes. This creates a sensitive issue about the amount of information that each voter can disclose without risking deanonymisation.

\section{Authority-minimized variants with an un-trusted source}
\label{Sec:authority}
Another approach towards anonymous quantum voting is to use a limited authority managing the resources and using an un-trusted source, and create more dependency on observable correlations \cite{xu2023decentralized, centrone2022quantum}. The device distributing the multipartite state can be un-trusted, and the correctness is inferred from test rounds and correlation statistics. This carries the ideas from device-independent quantum cryptography and integrates them with quantum voting schemes \cite{acin2007device, lo2012measurement}. If the collector is present, they can also be tested or avoided by switching to self-tallying. The steps are as follows,
\begin{enumerate}
    \item Randomly designate many rounds as tests and the rest as data.
    \item In test rounds, measure settings that check non-classical correlations and device behavior, and it can be aborted if errors are over the threshold limits.
    \item In data rounds, encode votes as usual. Use either a joint-measurement tally or the self-tallying combine.
    \item Publish correlation scores and the tally evidence.
\end{enumerate}
This scheme reduces the reliance on designated authorities and replaces it with public correlation tests that anyone can audit. The main issues with this scheme are the higher experimental demands, and in a small committee, very careful finite analysis is needed.

\section{Practical Implementation and Challenges}
\label{Sec:challenges}
Quantum voting is most feasible on a smaller scale, i.e., with small committees, keeping the current technology in mind. A more realistic implementation uses either entanglement-based sharing with simple local operations or a self-tallying design where each voter posts a compact share, and anyone recomputes the result \cite{wang2016self}. In either case, runs interleave data with test rounds to bound disturbance and loss, and all evidence is posted on an append-only bulletin board as in end-to-end verifiable elections.

Engineering challenges are urgent. Channel loss, alignment drift, and background light degrade quantum signals. Possible mitigation strategies include short links, active alignment, and narrow spectral or spatial filtering, with conservative abort thresholds derived from parameter estimation. Detector non-idealities and side channels require characterization and shielding. Finite-size effects are pronounced in small electorates, so the collector (the initiator) should allocate sufficient test rounds and report confidence intervals rather than point estimates. Fairness and simultaneous posting are enforced procedurally using commit-then-open or fixed windows to prevent adaptation after observing others. Finally, practical deployments must publish an assumption checklist that covers device calibration bounds, authenticated channels, sampling rates, dispute procedures, and explicit abort conditions. Within these constraints, self-tallying or memory-free optical designs are credible for committee-scale projects. However, larger deployments will depend on progress in scalable entanglement distribution and device certification.

\section{Advantages and Disadvantages}
\label{Sec:adv}
Since the different quantum voting schemes are explained, we discuss the advantages and disadvantages of these schemes over their classical counterparts. While post-quantum cryptography hardens classical systems against future quantum attacks, it still depends on the mathematical-hardness assumptions, which might not be enough at the current research stage. Quantum schemes allow for tampering detection by introducing noise in the measurement; however, these are tricky for practical implementations. 

\subsection{Advantages}
The advantages of quantum voting schemes over classical ones are as follows,
\begin{enumerate}
    \item Tamper detection due to elevated error rates in measurements when eavesdropping happens. 
    \item No cloning principle prevents no votes can be copied, and thus each voter can only vote once. 
    \item Entanglement supports anonymity by keeping local shares anonymous while revealing only the tally.
    \item Self-tallying designs let anyone recompute the result from posted data, which simplifies audits.
\end{enumerate}
\subsection{Disadvantages}
While quantum voting schemes have several advantages over their classical counterparts, there are several disadvantages to the large-scale deployment of quantum hardware. Some of them are as follows,
\begin{enumerate}
    \item Sensitive to loss, noise, alignment, and device side channels. It also requires calibration bounds and additional authenticated classical channels.
    \item Have high experimental overhead, such as entanglement sources, stable phase references, or joint measurements are demanding outside labs.
    \item Throughput and scale are quite limited today, as finite samples can make statistical checks noisy.
\end{enumerate}

Quantum voting schemes have several advantages over classical voting schemes, especially in terms of tamper-resistant and secure result declaration. However, due to the fragile nature of quantum hardware, the practical deployments face several challenges. The most prominent one is the noise in the hardware, which affects the transmission of qubits over longer distances, thus severely affecting the scalability of such setups. Quantum voting schemes can have several potential applications, with the ever-growing research studies to make quantum systems more stable and scalable in the long run.  

\section{Conclusion}
\label{Sec:Conclusion}

Quantum voting schemes provide information-theoretic security as promised by quantum mechanics principles; however, there are several challenges to the practical application of such schemes. This article goes over the basic mathematics of quantum voting schemes and then discusses the family of quantum voting schemes, including entanglement-based designs with a central tally, self-tallying protocols, and authority-minimized approaches that certify devices by correlations. Pilot-scale deployments have been explored in laboratories and small committees, and there is growing interest from standards bodies and consortia that already use end-to-end verifiable procedures. The central challenges are practical: distributing high-quality states over short but imperfect links, maintaining alignment and phase stability, characterizing detectors and side channels, authenticating all classical communication, and integrating eligibility and dispute handling without leaking ballots. These constraints severely limit the scaling of quantum voting schemes.

Moving forward, a realistic path combines quantum mechanisms for tamper-evident distribution and anonymity with classical end-to-end practices and post-quantum cryptography for authentication and record keeping. This deployment includes developing more reliable multipartite state distribution, noise and loss-tolerant protocol variants, measurement-device-independent or device-independent certification, and finite-size analyses. These are some of the challenges that need to be overcome to make quantum voting schemes a realistic, practical deployment for a large set of voters. However, in the near-term, quantum voting serves mainly for a small-scale scale well-controlled group.

\bibliographystyle{ACM-Reference-Format}
\bibliography{reference}

\end{document}